\documentstyle[11pt,epsfig]{article}
\input{epsf}
\begin{document}
\setlength{\baselineskip}{0.18in}
\newcommand{\be}{\begin{eqnarray}}
\newcommand{\ee}{\end{eqnarray}}
\newcommand{\bi}{\bibitem}
\newcommand{\lar}{\leftarrow}
\newcommand{\rar}{\rightarrow}
\newcommand{\lrar}{\leftrightarrow}
\newcommand{\hee}{{\cal H}_{ee}}
\newcommand{\hmm}{{\cal H}_{\mu\mu}}
\newcommand{\hem}{{\cal H}_{e\mu}}
\newcommand{\hes}{{\cal H}_{es}}
\newcommand{\hms}{{\cal H}_{\mu s}}
\newcommand{\hmt}{{\cal H}_{\mu \tau}}
\newcommand{\het}{{\cal H}_{e \tau}}
\newcommand{\hts}{{\cal H}_{\tau s}}

\newcommand{\dms}{\Delta_{\mu s}}
\newcommand{\dme}{\Delta_{\mu e}}
\newcommand{\dmt}{\Delta_{\mu \tau}}
\newcommand{\des}{\Delta_{e s}}

\newcommand{\deta}{\Delta_{e \tau}}
\newcommand{\dts}{\Delta_{\tau s}}
\newcommand{\dsm}{\Delta_{s\mu }}
\newcommand{\dem}{\Delta_{e\mu}}
\newcommand{\dtm}{\Delta_{\tau \mu}}
\newcommand{\dse}{\Delta_{se}}
\newcommand{\dte}{\Delta_{\tau e}}
\newcommand{\dst}{\Delta_{s\tau }}

\newcommand{\gee}{\gamma_{ee}}
\newcommand{\gem}{\gamma_{\mu\mu}}
\newcommand{\ges}{\gamma_{es}}
\newcommand{\gms}{\gamma_{\mu s}}
\newcommand{\rem}{\rho_{e\mu}}
\newcommand{\rme}{\rho_{\mu e}}
\newcommand{\res}{\rho_{es}}
\newcommand{\rse}{\rho_{se}}
\newcommand{\rms}{\rho_{\mu s}}
\newcommand{\rsm}{\rho_{s\mu}}
\newcommand{\rmm}{\rho_{\mu \mu}}
\newcommand{\ree}{\rho_{ee}}
\begin{center}
\vglue .06in
{\Large \bf { 
Can modified gravity explain accelerated cosmic expansion?
  }
}
\bigskip
\\{\bf A.D. Dolgov}$^{(a)(b)(c)}$ and 
{\bf M. Kawasaki}$^{(c)}$
\\[.2in]
$^{(a)}${\it INFN, sezione di Ferrara,
Via Paradiso, 12 - 44100 Ferrara,
Italy} \\
$^{(b)}${\it ITEP, Bol. Cheremushkinskaya 25, Moscow 113259, Russia.
}  \\
$^{(c)}${\it Research Center for the Early Universe, Graduate School of 
Science, \\University of Tokyo, Tokyo 113-0033, Japan
}

\end{center}

\vspace{.3in}
\begin{abstract}

We show that the recently suggested explanations of cosmic acceleration
by the modification of gravity at small curvature suffer violent
instabilities and strongly disagree with the known properties of
gravitational interactions.

\end{abstract}

\bigskip

It seems established that at the present epoch the universe expands with
acceleration. It follows directly from the observation of high red-shift
supernovae~\cite{sn-high-z} and indirectly from the measurements of
angular fluctuations of cosmic microwave background fluctuations
(CMBR)~\cite{cmbr}. The latter shows that the total mass/energy density
of the universe is very close to the critical one ($\Omega =1$), while
the observations of the universe large scale structure indicate that
normal gravitating (but invisible) matter can contribute only 30\% into
the total one. Thus one concludes that the remaining 70\% is some
mysterious agent that creates the cosmological acceleration. The
simplest suggestion is that the source of this acceleration is the
vacuum energy (cosmological constant). However such an explanation
immediately meets two serious problems. First, vacuum energy remains
constant in the course of cosmic expansion (modulo possible phase
transitions which could discontinuously change the value of $\rho_{vac}$
by a very large amount). On the other hand, ``normal" cosmological energy
density scales with time as $\rho_m\sim m^2_{Pl}/t^2$ and it is quite
strange coincidence that $\rho_{vac}$ and $\rho_m$ are of the same order
of magnitude just today. Second and much more serious, there are known
contributions into vacuum energy which are 50-100 orders of magnitude
larger than the cosmologically allowed value. At the present day there
is no workable scenario to solve the second problem, while the first one
may be solved if instead of a constant $\rho_{vac}$ a new massless or
very light field is introduced which evolves in the course of the
expansion in such a way that its energy density today more or less
naturally approaches the energy density of normal 
matter~\cite{Quintessense}.

Still such field remains mysterious. It would be in much better shape if
it simultaneously solved the problem of almost complete elimination
of gigantic contributions into vacuum energy by some adjustment
mechanism but, alas, it does not happen or, better to say, it is unknown
how it can be realized. To avoid an introduction of additional fields
(plurality should not be posited without necessity", W. Ockham) it was
suggested recently that gravity itself, if properly modified, could
create cosmological acceleration~\cite{capozziello}. To this end 
the model with the following action was considered: 
\be {\cal A}=\int d^4x \sqrt{-g}
\left[f(R)+{{\cal L}}_{M} \right]\,{,} \label{action} 
\ee 
where $R$ is the scalar curvature and ${\cal L}_M$ is the matter
Lagrangian. In the usual Einstein gravity the function $f(R)$ has the
form $f(R) = R m_{Pl}^2/16\pi$ where $m_{Pl}$ is the Planck mass
($\simeq 10^{19}$GeV). In ref.~\cite{capozziello} a power law
modification $f(R) \sim R^n$ with $n=-1, 3/2$ was considered.

A detailed examination of the $n=-1$ case was performed in the recent
paper~\cite{carrol}. The action was taken in the form: 
\be 
{\cal A} = \frac{m_{Pl}^2}{16\pi}\int d^4 x\,
\sqrt{-g}\left(R-\frac{\mu^4}{R}\right) 
-\int d^4 x\, \sqrt{-g}\, {\cal L}_M \, , 
\ee
and the corresponding equation of motion reads: 
\be
\left(1+\frac{\mu^4}{R^2}\right)R_{\alpha\beta} -
\frac{1}{2}\left(1-\frac{\mu^4}{R^2}\right)Rg_{\alpha\beta} +
\mu^4g_{\alpha\beta} \nabla_{\nu}\nabla^{\nu}\left(\frac{1}{R^2}\right)
-\mu^4\nabla_{(\alpha}\nabla_{\beta
)}\left(\frac{1}{R^2}\right)
=\frac{8\pi\,T_{\alpha\beta}^M}{m_{Pl}^2} \, ,\label{drmunu} 
\ee
where $T_{\mu\nu}^M$ is the matter energy-momentum tensor.  As argued in
ref.~\cite{carrol}, with the parameter $\mu$ chosen of the order of the
inverse universe age, $\mu^{-1} \sim 10^{18} \sec \sim (10^{-33}{\rm
eV})^{-1}$, this equation, applied to cosmology, describes the universe
acceleration in the present epoch, while the additional terms were
not essential at earlier times.

However an addition of non-linear in curvature terms into the action
integral is not an innocent step. It gives rise to higher order
equations of motion which possess quite unpleasant pathological
behavior. They may break unitarity and lead to ghosts or tachyons. And
if breaking of the usual gravity at high curvatures may be tolerated -
who knows what happens there - but breaking of gravity at low curvature
in weak field regime immediately leads to contradiction with well
established facts.  To demonstrate that let us consider equation of
motion for curvature scalar. It can be obtained from eq.~(\ref{drmunu})
by taking trace over $\alpha$ and $\beta$:
\be
D^2 R - 3 \,\frac{(D_\alpha R)\,(D^\alpha R)}{R} +\frac{R^4}{6\mu^4} 
- \frac{R^2}{2} = -\frac {T\,R^3 }{6\mu^4}\,,
\label{dr}
\ee
where $D$ is the covariant derivative and $T=8\pi T_\nu^\nu/m_{Pl}^2>0$.

Let us apply this equation for the gravitational field of a normal
celestial body, as e.g. the Earth or the Sun or any other smaller
gravitating object with weak gravity. In this case the metric can be 
approximately taken 
as a flat one, so $D^2 = \partial_t^2 -\Delta$ and $(D_\alpha
R)(D^\alpha R)= (\partial_t R)^2 - (\partial_j R)^2$. We will look for a
perturbative solution inside some spatially finite matter
distribution. In the lowest order the curvature is algebraically
expressed through the trace of the energy-momentum tensor of matter,
$R_0 = -T$. This is the standard result of General Relativity
(GR). Outside matter the solution is $R_0=0$ but now this is not exact
but true only in the lowest order approximation. One can check both
numerically and analytically that stationary solution outside matter
sphere is very quickly vanishing if $\mu^4 >0$. So at least in this case
the curvature scalar behaves similarly to the usual one. This is in
agreement with the analysis of stationary solutions of ref.~\cite{dick}
where it is argued that the modified gravity theories agree with the
Newtonian limit of the standard gravity for sufficiently 
small $\mu$.\footnote{
In Ref.~\cite{dick}, however, it was found that the condition for 
the correct Newtonian limit is very marginally satisfied for 
the simple $1/R$ gravity.
Thus, it may be likely that the deviation from the Newtonian gravity
is observed even in vacuum solutions. }

However for negative $\mu^4$ the solution is rising outside the matter
sphere and strongly deviates from the standard General
Relativity. Even for $\mu^4>0$, the curvature scalar, $R$,
outside the source would be non-vanishing and proportional to the
curvature on the boundary of the gravitating body but, as we mentioned
above, it tends to zero at very small distances from the boundary.

Let us consider now the internal solution with time dependent matter
density. The first order correction to the curvature, $R=-T+R_1$
satisfies the equation:
\be
\ddot R_1 &-&\Delta R_1 - \frac{6\dot T}{T}\, \dot R_1 
+\frac{6\partial_j T}{T}\, \partial_j R_1 + 
 R_1 \left[ T + 3\,\frac{\dot T^2 - (\partial_j T)^2}{T^2}  
- \frac{T^3}{6\mu^4} \right] \\ \nonumber
& =& D^2 T + \frac{T^2}{2} - \frac{3 D_\alpha T\,D^\alpha T}{T}\, ,
\label{ddotr1}
\ee
where the deviation of the metric from the Minkowski one may be
accounted for in the r.h.s.  by the covariant derivatives (but this is
not essential for our conclusion).

The last term in the l.h.s. has a huge coefficient: 
\be
T^3/6\mu^4 \sim 
(10^{-26}k^2 {\rm sec})^{-2}  
\left(\frac{\rho_m}{{\rm g\ cm}^{-3}}\right)^3\, ,
\label{t-eff}
\ee
where $\rho_m$ is the mass density of the gravitating body, the constant
$k$ is assumed to be of the order of unity, $\mu = k/t_u$ and $t_u
\approx 3\cdot 10^{17}$ sec is the cosmological time scale. 
This coefficient is much larger than $T$ since 
\be
T \sim (10^{3} {\rm sec})^{-2}
\left(\frac{\rho_m}{{\rm g\ cm}^{-3}}\right) .
\ee
Thus, the $T^3/6\mu^4$ dominates the coefficient in front of 
$R_1$ term in Eq. (\ref{ddotr1})
and  leads to strong instability.  (Notice that $-T^3/6\mu^4 < 0$.) 
The characteristic time of the instability rise is about $10^{-26}$
sec. Perturbation with the wave length larger than one tenth of
the proton Compton
wave length should be unstable. This result shows that the curvature
inside matter sphere should quickly rise and reach very high values 
(strong gravity) till some non-perturbative effects would terminate this
rise. This is the same kind of instability that leads to the universe
acceleration but now it is very unfavorable for the model and forbids
such a modification of gravity.

It was argued in a recent work~\cite{nojiri} that the considered above
modification of gravity follows from certain compactification schemes of
string/M-theory. If so, such theories must be excluded because of
the instability found in the present paper.

The analysis performed above is valid for a particular case of $1/R$
terms in the action integral.  A more general Lagrangian with
arbitrary powers of $R$, both negative and positive, is presented in the
recent work~\cite{capo2}. We have not made complete analysis of all
possible cases discussed in the literature but it seems suggestive that
all modifications of gravity at small curvature described by higher
order differential equations may suffer from serious instability
problems.
 
There are other earlier works on modifications of
gravity~\cite{mod-grav-1}-\cite{mod-grav-n} which might lead to
accelerated cosmological expansion.  However it is impossible to apply
the similar analysis to some of them because the theory is not
sufficiently specified - only a modification of the Friedmann equations
is suggested but no modified action which would allow to derive
equations in general case was
presented~\cite{mod-grav-1}-\cite{dvali03}. Another group of works is
based on modification of gravity due to higher
dimensions~\cite{damour02}-\cite{mod-grav-n}.  Probably these models do
not manifest instability discussed here, though only cosmological
solutions have been considered in the quoted papers. In
ref.~\cite{lue03} it is argued that modification of gravity at large
scales to induce cosmic acceleration can lead to a modification of
gravitational interactions at smaller scales but since the authors did
not specify their model it is hard to make any definitive conclusions.

To summarize, we have shown that an attempt to explain the observed
cosmic acceleration by the concrete modification of gravitational action
adding to the latter the term $\sim 1/R$ is unacceptable because
such term would lead to a strong temporal instability resulting in a
dramatic change of gravitational fields of any gravitating bodies. We
cannot make any conclusion about some other versions of the modified
gravity theory which could lead to cosmic acceleration because the latter 
are not sufficiently well defined to permit such an analysis.

\bigskip

{\bf Acknowledgment} A.D. Dolgov is grateful to the Research Center for
the Early Universe of the University of Tokyo for the hospitality during
the time when this work was done.

\end{document}